\documentclass[11pt]{article}

\usepackage{multirow}

\def\beq{\begin{eqnarray}}
\def\eeq{\end{eqnarray}}

\def\al{\alpha}
\def\be{\beta}

\def\ga{\gamma}

\def\vp{\varepsilon}

\def\la{\lambda}
\def\na{\nabla}
\def\pa{\partial}

\def\si{\sigma}

\def\Ga{\Gamma}

\begin{document}

\begin{center}
{\large\sc 
On the exact Foldy-Wouthuysen transformation for a Dirac spinor
\\
in torsion and other CPT and Lorentz violating backgrounds}
\vskip 8mm

%%%%%%%%%%%%%%%%%%%%%%%%%%%%%%%%%%%%%%%%%%%%%%%%%%%%%%%%%%%%
Bruno Gon\c{c}alves$^{a}$,
\quad
Yuri N. Obukhov$^{b}\footnote{On leave from: Dept. of Theoret.
Physics, Moscow State University, 117234 Moscow, Russia}$,
\quad
Ilya L. Shapiro$^{a}\footnote{Also at Tomsk State Pedagogical
University, Tomsk, Russia. \ E-mail address: shapiro@fisica.ufjf.br}$
\vskip 8mm

(a) \ \ {\small\sl Departamento de F\'{\i}sica, ICE,
Universidade Federal de Juiz de Fora
\\
Juiz de Fora, CEP: 36036-330, MG,  Brazil}
\vskip 5mm

(b) \ \ {\small\sl
%% Institute for Theoretical Physics, University of Cologne,
%% \\ 50923 K\"oln, Germany
Department of Mathematics, University College London
\\
Gower Street, London WC1E 6BT, UK
}

\end{center}
%%%%%%%%%%%%%%%%%%%%%%%%%%%%%%%%%%%%%%%%%%%%%%%%%%%%%%%%%%%%%%%%%%
\vskip 10mm

\begin{quotation}
\noindent
{\large\bf Abstract.}
We discuss the possibility to perform and use the exact
Foldy-Wouthuysen transformation (EFWT) for the Dirac spinor
coupled to different CPT and Lorentz violating terms. The
classification of such terms is performed, selecting
those of them which admit EFWT. For the particular
example of an axial vector field, which can be associated
with the completely antisymmetric torsion, we construct an
explicit EFWT in the case when only a timelike component of
this axial vector is present. In the cases when EFWT is
not possible, one can still use the corresponding technique
for deriving the perturbative Foldy-Wouthuysen
transformation, as is illustrated in a particular
example in the Appendix.

\vskip 4mm

\noindent
{\sl Keywords:} \
Foldy-Wouthuysen transformation,
CPT and Lorentz violating terms,
Torsion,
Magnetic Field.
\vskip 2mm

\noindent
{\sl PACS:}   %% %% %%  12.90.+b; 04.62.+v; 11.15.Kc
\ 12.90.+b; \ %%    Miscellaneous theoretical ideas and models
\ 04.62.+v; \ %% Quantum field theory in curved spacetime.
11.15.Kc.     %% Classical and semiclassical techniques
\vskip 2mm

\noindent
{\sl MSC-AMS:}
\ 81T10; \ %% Model quantum field theories
\ 83D05; \ %% Relativistic gravitational theories other than
           %% Einstein's, including asymmetric field theories
81T20.     %% Quantum field theory on curved space backgrounds.

\end{quotation}
%%%%%%%%%%%%%%%%%%%%%%%%%%%%%%%%%%%%%%%%%%%%%%%%%%%%%%%%%%%%%%
\vskip 12mm

% \newpage
%%%%%%%%%%%%%%%%%%%%%%%%%%%%%%%%%%%%%%%%%%%%%%%%%%%%%%%%%%%%%
%%%%%%%%%%%%%%%%%%%%%%%%%%%%%%%%%%%%%%%%%%%%%%%%%%%%%%%%%%%%%
%%%%%%%%%%%%%%%%%%%%%%%%%%%%%%%%%%%%%%%%%%%%%%%%%%%%%%%%%%%%%
\section{Introduction}

One of the most natural extensions of General Relativity is
related to the inclusion of the spacetime torsion which is
supposed to describe, along with the metric, the physical
properties of the spacetime geometry. The study of the physical
aspects of the torsion gravity has a long history (see
\cite{Hehl,IS,PBO,Gas,book,torsi} for extensive reviews and
references). The issue which always attracted a special attention
was the interaction of the spacetime torsion with the
spinor field and with the spinning particle \cite{D,A,H,R}.
In particular, the papers \cite{Pronin,BBS,rytor} were devoted
to the nonrelativistic approximation of Dirac equation and
in \cite{BBS,rytor}, correspondingly, the Pauli equation and
Foldy-Wouthuysen transformation have been obtained for the
fermion field coupled to the combined electromagnetic and torsion
fields. One can use these results for the investigation of the
possible manifestations of torsion in the domain of the atomic
physics \cite{BBS,lamme,KhZhu}.

The Foldy-Wouthuysen transformation provides, in general,
more detailed information about the nonrelativistic
approximation \cite{FW}, especially if the exact version of this
transformation is constructed \cite{EK,nikitin,CASE,OT,diraceq}
(see also recent works \cite{SiTer}).
It is, in principle, safer to perform the exact transformation,
since otherwise there is a certain risk of missing some important
terms. Recently it has been shown that this is the case for the
spinor field in the weak gravitational field \cite{diraceq}.
Therefore it is worthwhile to construct EFWT for the case of
torsion and electromagnetic background.
One can imagine, for instance, the situation when the magnetic
field could amplify the effect of torsion and thus make the
upper bound for the torsion more precise. Recently, we have used
this approach for the case of a fermion on the combined background
of the gravitational wave and magnetic field and found that
indeed there are potentially interesting nonlinear effects
\cite{Gon}. In the present paper we mainly
consider the case of torsion. In fact, the same approach can
be used also for other Lorentz and CPT violating terms
\cite{KostPott,CollKost}. Although the main aim of our work
is to study the torsion effects, in section \ref{VT} we
present a table which shows the possible CPT and Lorentz
violation terms that could be treated using this technique.

The usual perturbative Foldy-Wouthuysen transformation can be
constructed for the Dirac field interacting with the variety
of external fields, including the torsion \cite{rytor}. However,
the possibility to have an exact FW transformation depends on
a special condition (the existence of the involution operator)
on the external classical fields. The construction of an exact
transformation is more complicated and more interesting from the
mathematical point of view \cite{EK,nikitin}. As it was already
mentioned above, in this paper we are interested in the set of two
external fields - one is the torsion and another one is a constant
and uniform magnetic field. One can safely assume that torsion is
very weak, since otherwise it would be easy to detect \cite{torsi},
while the magnetic field of our interest should be very strong.
Therefore, our goal should be to find the transformation which is
exact in magnetic field but may be just linear in torsion. Actually,
EFWT with a general torsion is not always possible because the
corresponding Hamiltonian has a term that does not admit the
involution operator. So we construct the transformation using only
the scalar part of the torsion field to describe it.

In Appendix, we take into account the vector part of
the torsion, introducing some {\it ad hoc} modification
of the torsion-dependent term in the Hamiltonian, such that the
modified expression admits the involution operator. Then one
can use the known technique developed for EFWT. The main point
is that, in the linear approximation, the mentioned modification
can be easily removed from the final result. In this way we
can reproduce the known perturbative result \cite{rytor} in
a technically much more economic way and also to get the 
Foldy-Wouthuysen Hamiltonian with the terms which show
explicitly the mixture between the torsion and magnetic field.
In other words, we have derived a Hamiltonian which is exact in
magnetic field and linear in torsion.
After performing the transformation we derive
the non-relativistic equations of motion for the particle with
spin $\frac{1}{2}$.

The paper is organized as follows. In the next section we
study the possibility to apply EFWT to the CPT and Lorentz
violating terms. In section \ref{FWT} we consider an example
of EFWT in the torsion case. In section \ref{con}, we draw our
conclusions, and in Appendix we discuss the linear
expansion in the torsion field. Throughout the paper
we use Greek letters for the indices which run from
$0$ to $3$. Latin indices are used for the space
coordinates and run from $1$ to $3$.

%%%%%%%%%%%%%%%%%%%%%%%%%%%%%%%%%%%%%%%%%%%%%%%%%%%%%%%%%%%%%
%%%%%%%%%%%%%%%%%%%%%%%%%%%%%%%%%%%%%%%%%%%%%%%%%%%%%%%%%%%%%
%%%%%%%%%%%%%%%%%%%%%%%%%%%%%%%%%%%%%%%%%%%%%%%%%%%%%%%%%%%%%
\section{EFWT for Dirac equation with CPT and Lorentz Violating
Terms}
\label{VT}

Let us start with the action describing a Dirac fermion
with Lorentz and CPT symmetry breaking terms. For the sake
of generality, we include also minimal interaction to
gravity.
\beq
S & = & \int d^4x\sqrt{-g}\,\left\{\,
\frac{i}{2}\,\bar{\psi}\,\Ga^\mu D_\mu\psi
\,-\, \frac{i}{2}\, D^{\star}_\mu \bar{\psi}\,\Ga^\mu\psi
\,-\,\bar{\psi}\, M\,\psi \,\right\}\,,
\label{action0}
\eeq
where we use the following classification for the
possible Lorentz and CPT symmetry breaking terms
\cite{CollKost}
\beq
D_\mu = \na_\mu - i\,e\,A_\mu\, ; \;\;\; D^{\star}_\mu
= \na_\mu + i\,e\,A_\mu\, ; \;\;\;
\Ga^\nu = \ga^\nu + \Ga_1^\nu\, ; \;\;\; M = m + M_1\,.
\eeq
Here $\na_\mu$ is the operator of the covariant derivative,
$F_{\mu\nu} = \na_\mu A_\nu - \na_\nu A_\mu$ and the
quantities $\Ga_1^\nu$ and $M_1$ are given by
\beq
\Ga_1^\nu & = & c^{\mu\nu}\ga_\mu + d^{\mu\nu} \ga_5\ga_\mu
+ e^\nu + i\,f^\nu\ga_5
+\frac{1}{2}\,g^{\la\mu\nu}\si_{\la\mu}\, ,
\label{break1}
\\
M_1 & = & a_\mu\,\ga^\mu + b_\mu\,\ga_5\,\ga^\mu
+ i\, m_5\ga_5 + \frac{1}{2}\, H_{\mu\nu}\,\si^{\mu\nu}\, .
\label{break2}
\eeq
The quantities
$a_\mu$, $b_\mu$, $m_5$, $c^{\mu\nu}$, $d^{\mu\nu}$, $e^\mu$,
$f^\mu$,
$g^{\la\mu\nu}$ and $H_{\mu\nu}$ are CPT and/or Lorentz violating
parameters.
An extensive discussion of the possible origin of these parameters
and also their numerous phenomenological implications can be found
in \cite{LCPT1,LCPT2} and we will not consider these aspects
here. From now on we are going to treat these terms as constants,
so it is possible to rewrite (\ref{action0}) in the following
way
\beq
S & = & \int d^4x\sqrt{-g}\,\left\{\,i
\bar{\psi}\,\Ga^\mu D_\mu\psi
-\bar{\psi}\, M\,\psi \,\right\}\,.
\label{action01}
\eeq

As a result, the equations of motion for $\psi$ can be written
as $i\Ga^\mu D_\mu\psi\,=\,M\,\psi$. In order to perform
EFWT we put this equation into the Schr\"{o}dinger form,
$i\pa_t \psi \,=\, H \psi$, to get the Hamiltonian 
\beq
i\Ga^0 \na_0 \psi = (M + \Ga^\mu P_{\mu}^*) \psi\,.
\label{eqpsi}
\eeq
Here we introduced the useful notations
\beq
\overline{P}_{\nu} = (0,P_i)
\qquad \mbox{and} \qquad
P_{\nu}^* = \overline{P}_{\nu} - e A_{\nu}
\eeq
and use the standard representation for the Dirac matrices
(see, for example, \cite{BD})
\beq
\beta &=& \gamma^0 = \left(\matrix{1 &0\cr 0 &-1\cr} \right)
\,,\qquad
\alpha_i = \be \gamma_i =
\left(\matrix{0 &\sigma_i\cr \sigma_i & 0\cr} \right)\,,
\nonumber
\\
\gamma_5 &=& i\gamma^0 \gamma^1 \gamma^2 \gamma^3
\,,\qquad\qquad\quad
\sigma_{\mu \; \nu} = \frac{i}{2}(\gamma_\mu \gamma_\nu - \gamma_\nu \gamma_\mu)\,.
\label{matrices}
\eeq

Let us denote $\Ga^0 = \ga^0 + \Ga^0_1$ and introduce
$\overline{\Ga}^0_1$ such that
$(\Ga^0)^{-1} = \ga^0 - \overline{\Ga}^0_1$.
If one assume that the Hamiltonian is linear in the
CPT/Lorentz violating terms present in $\Ga^0_1$, it is
straightforward to check that
$$
\overline{\Ga}^0_1 = \ga^0 \, \Ga^0_1 \, \ga^0\,.
$$
Therefore, the equation (\ref{eqpsi}) can be recast into the
following form: 
\beq
i\na_0 \psi \, = \,
\Big[\ga_0 \,-\, \ga_0(c^{\mu 0}\ga_\mu +
d^{\mu 0} \ga_5\ga_\mu
+ e^0 + i\,f^0\ga_5
+\frac{1}{2}\,g^{\la\mu 0}\si_{\la\mu})\ga_0 \Big]\times
\big( M + \Ga^\nu P^*_\nu \big)\psi \,.
\label{ham}
\eeq

It is possible to construct an exact FW transformation if
\beq
JH + HJ = 0
\,\,, \qquad
\mbox{where} \qquad
J = i\gamma_5\beta
\label{involution}
\eeq
is the involution operator. Only those theories where the
Hamiltonian admits the involution operator enable one to
perform EFWT \cite{EK,nikitin,CASE,diraceq}. One thus can
formulate the natural question: Which is the most general
form of equation (\ref{ham}) that admits the involution
operator? In order to answer this question, one has to check
whether the criterion (\ref{involution}) is satisfied for
the terms in the general Hamiltonian presented above in
the right hand side of (\ref{ham}). The result of this
procedure is given in the Table.

\begin{table}[htb]
   \centering
   \large      % tamanho da fonte
   \setlength{\arrayrulewidth}{1.5\arrayrulewidth}  % espessura da  linha
   \setlength{\belowcaptionskip}{10pt}  % espaço entre caption e tabela
   \caption{\it Interaction coefficients}
   \begin{tabular}{|c|c|c|c|c|c|c|c|c|}
      \hline
      \multirow{3}{0.2cm}{}
& $m$ & $a_l$ & $b_0$ & $H^{lj}$ & $m_5$& $b_l$& $a_0$& $H^{0 \mu}$ \\
& $e^{\nu}P^*_\nu$ & $c^{l \nu}P^*_\nu$ & $d^{0\nu}P^*_\nu$
& $g^{lj\nu}P^*_\nu$ & $f^\nu P^*_\nu$& $d^{l\nu}P^*_\nu$& $c^{0\nu}P^*_\nu$
& $g^{0\mu\nu}P^*_\nu$ \\

& & $P^*_l$ & & & & & $P^*_0$& \\
\hline \hline
$\ga^0$ & $1$ & $\ga^l$& $-\ga^0\ga^5$& $\frac{1}{2}\sigma^{lj}$
& & & & \\
      \hline
$c^{00}$ & $-\ga^0$ & $-\al^l$& $\ga^5$& $-\frac{1}{2}\ga^0\sigma^{lj} $
& & & & \\
      \hline
$f^0$ & $i\ga^5$ & $i\ga^5\ga^l$& $i\ga^0$& $\frac{i}{2}\ga^5 \sigma^{lj}$
& & & &\\
      \hline
$d^{i0}$ & $-i\ga^i\ga^5$&  $-i\ga^i\ga^5\ga^l$& $\al^i$&
$-\frac{1}{2}\ga^i\ga^5 \sigma^{lj}$& & & &\\
      \hline
$g^{i00}$ & $2\al^i$ & $2\al^i\ga^l$& $2\ga^i\ga^5$& $\al^i \sigma^{lj} $
& & & & \\
      \hline
$d^{00}$ & & & & & $i\ga^0$& $\al^l$& $-\ga^5$
& $\frac{1}{2}\sigma^{0 \mu}\ga^0\ga^5$\\
      \hline
$e^0$ & & & & & $-i\ga^5$& $-\ga^5\ga^l$& $-\ga^0$
& $-\frac{1}{2}\sigma^{0 \mu}$\\
      \hline
$c^{i0}$ & & & & & $-i\ga^i\ga^5$& $-i\ga^i\ga^5\ga^l$& $-\al^i$
& $\frac{1}{2}\ga^i\sigma^{0 \mu}$\\
      \hline
$g^{ij0}$ & & & & & $\frac{i}{2}\sigma^{ij}\ga^5$
& $\frac{1}{2}\sigma^{ij}\ga^5\ga^l$ & $\frac{1}{2}\sigma^{ij}\ga^0$
& $\frac{1}{4}\sigma^{ij}\sigma^{0 \mu}$\\
      \hline
   \end{tabular}
\end{table}

The Table specifies the 80 cases of CPT and Lorentz
violating terms in the modified Dirac equation which admit EFWT.
The form of the corresponding EFWT-positive term in the Hamiltonian,
is obtained by multiplying the terms in the row
and in the line. For example, the coefficient $1$ in the first row
and first column means that for $\ga^0$ and $m$ the Hamiltonian 
contains the term $\ga^0 \times m \times 1 = \be m$. Of course,
this term is the most trivial one as it corresponds to the free
Dirac equation.

Another example is for the 8-th line with  $c^{i0}$ and the 8-th
row with $g^{0\mu\nu}P^*_\nu$. Taking the coefficient inside the
Table into account, we arrive at the EFWT-admitting term
$$
c_{i0} \times g^{0\mu\nu}P^*_\nu \times
\frac{1}{2}\ga^i\sigma_{0 \mu} = \frac{i}{2}
\ga_ i c^{i0} \, \al_j g^{0 j \nu} \,  P^*_\nu\,.
$$
Each term in the zero line must be taken separately, e.g. in the
first row there are two different terms $m$ and $e^{\nu}P^*_\nu$.
The filled blocks with nonzero coefficients show the terms
which allow EFWT. As we have just mentioned, there are 80 such
terms which means the corresponding number of the modified Dirac
theories admitting EFWT.

Furthermore, if some space in the table is empty, this means
that EFWT is {\it not} allowed to the given pair of terms in
the corresponding row and line. The same is true if a component
of one term is not present on the table. Let us note that even
in those cases when the Hamiltonian does not satisfy the equation
(\ref{involution}), EFWT technique is not useless. In fact, 
there is a possibility to apply the EFWT prescription to
perform a perturbative Foldy-Wouthuysen transformation
and to achieve a reliable qualitative analysis of the even
transformed Hamiltonian. For the product of the terms $\ga^0$
and $P^*_0$ (which has an empty site in the table), the
corresponding calculation has been performed in \cite{Bruno}.
In Appendix, we analyze another interesting
example, namely the case of the product of $\ga^0$ and $b_l$
(space-like component of the axial vector, dual to the
completely antisymmetric torsion).

%%%%%%%%%%%%%%%%%%%%%%%%%%%%%%%%%%%%%%%%%%%%%%%%%%%%%%%%%%%%%
%%%%%%%%%%%%%%%%%%%%%%%%%%%%%%%%%%%%%%%%%%%%%%%%%%%%%%%%%%%%%
%%%%%%%%%%%%%%%%%%%%%%%%%%%%%%%%%%%%%%%%%%%%%%%%%%%%%%%%%%%%%
%% \section{Dirac field interacting with torsion}
%% \label{tor}

%%%%%%%%%%%%%%%%%%%%%%%%%%%%%%%%%%%%%%%%%%%%%%%%%%%%%%%%%%%%%
%%%%%%%%%%%%%%%%%%%%%%%%%%%%%%%%%%%%%%%%%%%%%%%%%%%%%%%%%%%%%
%%%%%%%%%%%%%%%%%%%%%%%%%%%%%%%%%%%%%%%%%%%%%%%%%%%%%%%%%%%%%
\section{Example of exact Foldy-Wouthuysen transformation}
\label{FWT}

In this section we consider in details the EFWT for one of
those cases which admit this exact transformation. Namely, we
construct the EFWT for the purely timelike axial vector field
which is dual to the completely antisymmetric torsion of the
spacetime. Let us start with some necessary details about the
gravity theory with torsion. We shall use the notations
of \cite{torsi}.

In the spacetime with torsion $T^\alpha_{\;\beta\gamma}$, the
connection
$\tilde{\Gamma}^\alpha_{\;\beta\gamma}$ is not symmetric,
$\tilde{\Gamma}^\alpha_{\;\beta\gamma} -
\tilde{\Gamma}^\alpha_{\;\gamma\beta} =
T^\alpha_{\;\beta\gamma}$.
It proves useful to divide torsion $T^\alpha_{\;\beta\gamma}$
into the following irreducible
components: the trace $T_{\beta} = T^\alpha_{\;\beta\alpha}$,
the pseudotrace
$\;S^{\nu} = \varepsilon^{\alpha\beta\mu\nu}T_{\alpha\beta\mu}\;$
and the pure tensor part
$q^\al_{\;\be\ga}$, satisfying the conditions
$q^\al_{\;\be\al} = \vp^{\al\be\mu\nu}q_{\al\be\mu}=0$.
Then torsion can be written in the form
\beq
T_{\alpha\beta\mu} = \frac{1}{3}\,\left( T_{\be}g_{\al\mu} -
T_{\mu}g_{\al\be} \right) - \frac{1}{6}\,\vp_{\al\be\mu\nu}
S^{\nu} + q_{\al\be\mu}\,.
\eeq
In what follows we shall consider only the $S_\mu$-component,
that is equivalent to taking completely antisymmetric torsion.

Since the Dirac fermion is in an external gravitational field
with the torsion, we can perform the minimal covariant
generalization of the flat-space action by replacing the
Minkowski metric by a general one and the partial derivative
by the covariant one. However it is somehow more interesting
to consider a general non-minimal action \cite{bush,book,torsi},
which includes all terms compatible with the covariance and with
no inverse-mass parameters,
\beq
S = \int d^4 x\,\sqrt{-g}
\Big\{i\bar{\psi}\ga^\mu(\na_\mu + i\eta_1\gamma_5
S_\mu)\psi+m\bar{\psi}\psi \Big\}\,.
\label{actiontor}
\eeq
Here $\eta_1=1/8$ corresponds to the minimal action case
\cite{bush}. According to \cite{bush,torsi} (see also
further references therein) the consistent quantum field
theory with the torsion can be constructed only for the
nonminimal interaction of Dirac field with the external
torsion field. Therefore in what follows we shall keep
the parameter $\eta_1$ arbitrary. Now if we put
(\ref{break2}) into (\ref{action0}) and compare
the result with (\ref{actiontor}), it is possible
to see that $b_\mu = -\eta_1 S_\mu$.

At this point, we are in a position to develop the
calculations of EFWT with one of the CPT/Lorentz violating
terms. Consider the spin-$1/2$ particle in an external
torsion and electromagnetic fields. We are going to consider
constant magnetic and torsion fields. The equation of motion
which follows from the action (\ref{actiontor}) has the form
\beq
i\hbar\,\frac{\pa\psi}{\pa t}\,=\,
\Big(c\overrightarrow{\alpha}\cdot\overrightarrow{p}-
e\overrightarrow{\alpha}\cdot\overrightarrow{A}-
\eta_1\overrightarrow{\alpha}\cdot\overrightarrow{S}\gamma_5+e\Phi+
\eta_1\gamma_{5} S_0+mc^2\beta\Big)\,\psi\,,
\label{ham0}
\eeq

In case of a constant magnetic field one can
set $\Phi=0$. However, a direct inspection shows that
the term $\eta_1\overrightarrow{\al}\overrightarrow{S}\ga_5$
in (\ref{ham0}) does not satisfy the condition (\ref{involution}). 
So let us first treat the case when $\overrightarrow{S}=0$
and EFWT can be derived. The complete Hamiltonian is studied
in Appendix. Here we work with the following Hamiltonian
\beq
H\,=\,
c\overrightarrow{\alpha}\cdot\overrightarrow{p}-
e\overrightarrow{\alpha}\cdot\overrightarrow{A}+
\eta_1\gamma_{5} S_0+mc^2\beta\,.
\label{ham1}
\eeq

An interesting point that has to be emphasized here is that the
above Hamiltonian  could have been constructed from the table
of section \ref{VT} without using the arguments of the
last two paragraphs. If we look to the $\ga^0$ line of the table
and we want to consider only the $b_\mu$ field, we conclude that
only the component $b_0$ is allowed. Therefore, the most general
Hamiltonian to torsion field using the table scheme would be
$\ga_0 \times (m + \ga^l P^*_l -\ga_0\ga^5 b_0)$,
that has the same form of (\ref{ham1}).

According to the standard prescription \cite{EK}, the next step
is to obtain $H^2$. Direct calculations yield the result
\beq
H^2 &=& (c\overrightarrow{p}-
e\overrightarrow{A}-
\eta_1\overrightarrow{\Sigma}S_0)^2+
m^2c^4 - 2\eta_1^2 S_0^2
\,\,.
\label{h2comp}
\eeq

In order to get the transformed Hamiltonian $H^{tr}$ we rewrite $H^2$
as $H^2=A^2+B$ with $A$ being $m$-dependent terms in $H^2$, whereas
the terms in $B$ do not depend on the mass. In the present case $A=mc^2$.
Then, we search for an operator $K$ in the form
\beq
K=A+\frac{1}{A}K_1+K_1\frac{1}{A}+\vartheta(\frac{1}{A^2})\,,
\eeq
such that $K^2=H^2$. Finally, using (\ref{h2comp}) and the fact that
\beq
H^{tr}\,=\,UHU^{*}\,=\,\beta[\sqrt{H^2}]^{EVEN}
+ J[\sqrt{H^2}]^{ODD}\,.
\label{tr}
\eeq
Here the even (odd) terms in (\ref{tr}) are the ones that
commute (anticommute) with the matrix $\beta$. We thus get
\beq
H^{tr} &=&
\beta mc^2+\frac{\beta}{2mc^2}(c\overrightarrow{p}
- e\overrightarrow{A}- \eta_1\overrightarrow{\Sigma}S_0)^2
-\beta\frac{\eta_1^2}{mc^2}\,S_0^2\,.
\eeq

The next step is to present the Dirac fermion field
$\psi$ in the form
\beq
\psi  =
\left( \begin{array}{c}
\varphi  \\
\chi    \\
       \end{array}
\right)
e^{\frac{-imc^2t}{\hbar}}\;\;,
\label{expo}
\eeq
and to use the equation $i\hbar\pa_t\psi=H\psi$ to derive
the Hamiltonian for the two-spinor $\varphi$. We obtain
the two-component equation
\beq
i\hbar\frac{\pa}{\pa t}
\left( \begin{array}{c} \varphi
\\
\chi   
\\
\end{array}
\right)
\,=\,\left(- mc^2+H\right)\,\,
\left( \begin{array}{c} \varphi
\\
\chi   
\\
\end{array}
\right)
\;.
\label{eqexpo}
\eeq
Using the fact that the transformed Hamiltonian
is an even function, we obtain, in the $\varphi$ sector, the
same nonrelativistic Hamiltonian of \cite{BBS}.

%%%%%%%%%%%%%%%%%%%%%%%%%%%%%%%%%%%%%%%%%%%%%%%%%%%%%%%%%%%%%
%%%%%%%%%%%%%%%%%%%%%%%%%%%%%%%%%%%%%%%%%%%%%%%%%%%%%%%%%%%%%
%%%%%%%%%%%%%%%%%%%%%%%%%%%%%%%%%%%%%%%%%%%%%%%%%%%%%%%%%%%%%
\section{Discussion and conclusion}
\label{con}

The new classification of the most general CPT and Lorentz
violating terms in the Dirac equation with respect to an exact
Foldy-Wouthuysen transformation was developed. We found
80 examples of the terms which admit such a transformation.

We have derived the exact Foldy-Wouthuysen
transformation for the Dirac spinor field on the combined
background of the torsion and the constant uniform magnetic
fields. We have constructed this for the fermion interacting
with the scalar part of torsion field $S_\mu$. Using the
method of \cite {diraceq, EK, nikitin, CASE} we were able to
reproduce known results \cite{BBS, rytor} in a much more
general and economic way. We also constructed a table which
gives the most general Hamiltonian for each of the CPT and
Lorentz violating terms that admit EFWT.

Although the vector part of torsion field does not admit
the exact transformation, in Appendix we present a
qualitative analysis of the term in the initial Hamiltonian
that allows for such a transformation. After a proper
modification of it, it is possible to find a transformed
Hamiltonian which is linear in torsion ($S_\mu$) and is
non-perturbative in the external constant magnetic field.
The same structure was obtained for the non-relativistic
equations of motion of a spinning particle. This qualitative
analysis demonstrates that in this case there is a mixing
between the magnetic and torsion fields terms. 
\bigskip

\noindent
{\large\bf Acknowledgments}.
The work of B.G. has been supported by FAPEMIG
(MG, Brazil) and by the PRONEX project from
FAPES (ES, Brazil).
The work of Yu.O. was partially supported by FAPESP
(S\~{a}o Paulo, Brazil) and by DFG (Bonn).
The work of I.Sh. has been supported by FAPEMIG
(MG, Brazil), CNPq (Brazil), by the PRONEX project from
FAPES (ES, Brazil) and by ICTP.
%%%%%%%%%%%%%%%%%%%%%%%%%%%%%%%%%%%%%%%%%%%%%%%%%%%%%%%%%%%
%%%%%%%%%%%%%%%%%%%%%%%%%%%%%%%%%%%%%%%%%%%%%%%%%%%%%%%%%%%%%%%%%%
%%%%%%%%%%%%%%%%%%%%%%%%%%%%%%%%%%%%%%%%%%%%%%%%%%%%%%%%%%%%%%%%%%

%%%%%%%%%%%%%%%%%%%%%%%%%%%%%%%%%%%%%%%%%%%%%%%%%%%%%%%%%%%%%
%%%%%%%%%%%%%%%%%%%%%%%%%%%%%%%%%%%%%%%%%%%%%%%%%%%%%%%%%%%%%
%%%%%%%%%%%%%%%%%%%%%%%%%%%%%%%%%%%%%%%%%%%%%%%%%%%%%%%%%%%%%
\newpage
\section{Appendix}

The Hamiltonian (\ref{ham1}) does not allow for an EFWT.
However, due to the weakness of the torsion field we are
really interested only in the linear order in torsion
while the magnetic field should be treated exactly.

Let us make an {\it ad hoc} modification of the term
$\eta_1\overrightarrow{\al}\overrightarrow{S}\ga_5$, that
is multiply it by the $\be$-matrix. The modified term
satisfies the condition (\ref{involution}) and now the
EFWT is perfectly
possible. The main point is that, in the linear order in the
torsion field, an extra $\be$ has no effect. The reason is that,
after deriving the final Hamiltonian operator, it will have the
block diagonal structure. We are interested only in the upper
block of Hamiltonian which is even (after transformation) to
perform the physical analysis. At least in the first order in
$1/m$, it does not matter if this term is multiplied by $\be$
or not, because beta has the form (\ref{matrices}) and its
upper block is just the unity matrix. As a result, we arrive
at what one can call semi-exact Foldy-Wouthuysen transformation,
because it is exact in only part of external fields and linear
in other external fields. This technique was already been
applied for a Dirac Hamiltonian including scalar electromagnetic
potential \cite{Bruno}.

For the sake of completeness we include also the timelike
component of the axial vector, $S_\mu$. After all, the
Hamiltonian we are going to deal with has the form
\beq
H\,=\,c\overrightarrow{\al}\cdot\overrightarrow{p}
- e\overrightarrow{\al}\cdot\overrightarrow{A}
- \,\eta_1\overrightarrow{\al}\cdot\overrightarrow{S}\ga_5\be
\,+ \eta_1\ga_{5} S_0+mc^2\be\,.
\label{H-ap}
\eeq
In this case $H^2$ has the form
\beq
H^2 &=& (c\overrightarrow{p}-
e\overrightarrow{A}-
\eta_1\overrightarrow{\Sigma}S_0)^2+
m^2c^4+2\eta_1mc^2\overrightarrow{\Sigma}\cdot\overrightarrow{S}
\nonumber
\\
&+& \eta_1^2(\overrightarrow{S})^2
+ \hbar ce\overrightarrow{\Sigma}
\cdot\overrightarrow{B}
- 2 \eta_1^2 S_0^2
+ 2i\eta_1\gamma_5\beta\overrightarrow{\Sigma}\cdot
\big[\overrightarrow{S}\times
(c\overrightarrow{p}-e\overrightarrow{A})\big]
\,\,.
\label{h2comp-ap}
\eeq
The last term in (\ref{h2comp-ap}) is odd, and its
presence looks somehow naturally, since we have used
the artificial procedure in (\ref{H-ap}). At the same
time, if we do drop this term, the rest is exactly the
Hamiltonian which follows from the usual perturbative
Foldy-Wouthuysen transformation with torsion \cite{rytor}.
An obvious advantage of the present method is its
technical simplicity compared to the perturbative one.

If we apply the procedure described between (\ref{h2comp})
and (\ref{eqexpo}) to the above equation, we find the
nonrelativistic limit which is almost (but not completely)
equal to the conventional one \cite{BBS},
\beq
H^{tr}_\varphi=\frac{1}{2m}(\overrightarrow{\Pi})^2+B_0+
\overrightarrow{\sigma}\cdot\overrightarrow{Q}\,,
\label{BBS}
\eeq
where
\beq
\overrightarrow{\Pi} &=& \overrightarrow{p}-
\frac{e}{c}\overrightarrow{A}-
\frac{\eta_1}{c} S_0 \overrightarrow{\sigma}
\,,\quad
B_0 = - \frac{\eta^2_1}{mc^2}\,S_0^2\,,
\nonumber
\\
\overrightarrow{Q} &=& \eta_1\overrightarrow{S}+
\frac{\hbar e}{2mc}\overrightarrow{B}
+\frac{\eta_1}{mc}
\overrightarrow{S}\times
\big(\overrightarrow{p}-\frac{e}{c}\,\overrightarrow{A}\big)\,.
\label{ham-ap}
\eeq
The very last term in $\overrightarrow{Q}$ originates from
the odd term in (\ref{h2comp}) which we already discussed above.
This term is new in comparison with the expressions derived
in \cite{BBS} and in \cite{rytor} through the usual perturbative
Foldy-Wouthuysen transformation. The fact that the exact
transformation gives a new term in comparison with the
perturbative transformation is analogous to the gravitational
case in \cite{diraceq}, described by the appearance of the
"gravitational Darwin" term.

The canonical quantization of (\ref{ham-ap}) gives us the
(quasi)classical equations of motion
\beq
\frac{dx_i}{dt}
= \frac{1}{m}
\left(p_i - \frac{e}{c}A_i - \frac{\eta_1}{c}\sigma_iS_0 \right)
+\frac{\eta_1}{mc}
\left[ \overrightarrow{\sigma}\times\overrightarrow{S} \right]_i
= v_i\,,
\nonumber
\eeq
\beq
\frac{dp_i}{dt}
= \frac{1}{m} \left(p^j - \frac{e}{c}A^j
- \frac{\eta_1}{c}\sigma^j S_0 \right)
\frac{e}{c}\frac{\pa A_j}{\pa x^i}
+\frac{\eta_1}{mc}
\left[ \overrightarrow{\sigma}\times\overrightarrow{S} \right]^j
\frac{e}{c}\frac{\pa A_j}{\pa x^i}
\,,
\nonumber
\eeq
\beq
\frac{d\sigma_i}{dt}
= \left[ \overrightarrow{R}\times\overrightarrow{\sigma} \right]_i
\quad,\quad
\overrightarrow{R}
= \frac{2\eta_1}{\hbar}\left[ \overrightarrow{S}
- \frac{1}{c}\overrightarrow{v}S_0
+\overrightarrow{S}\times\frac{\overrightarrow{v}}{c}
+\frac{2\eta_1}{\hbar}S_0
\overrightarrow{S}\times\overrightarrow{\sigma}\right]
+ \frac{e}{mc}\overrightarrow{B}\,.
\label{eqmotion-ap}
\eeq
The last equations are very similar to the ones derived
previously in \cite{BBS} and \cite{rytor} on the basis of
Pauli equation and perturbative Foldy-Wouthuysen transformation.
At the same time there are some extra terms due to the nonlinear
approximation in the external fields which we use here.

The first two equations of (\ref{eqmotion-ap}) give
\beq
m\frac{dv_i}{dt} = -\frac{e}{c}\frac{\pa A_i}{\pa t}
+ \frac{e}{c}\left[ \overrightarrow{v}
\times\overrightarrow{B} \right]_i
- \frac{\eta_1}{c}\sigma_i\frac{\partial S_0\sigma_i}{\partial t}
-\frac{\eta_1}{c}\frac{\partial
(\overrightarrow{S}\times\overrightarrow{\sigma})_i}{\partial t}\,.
\label{eqmotion2-ap}
\eeq

Now we can rewrite the equation (\ref{h2comp-ap}) using
the linear approximation in $S_\mu$. From now on, all the
terms that have power greater than two in $S_\mu$ will
be neglected. We find
\beq
H^2=H_0^2+2\eta_1mc^2
\overrightarrow{\Sigma}\cdot\overrightarrow{S}+
2\eta_1\gamma_5S_0
\overrightarrow{\alpha}\cdot
(c\overrightarrow{p}-e\overrightarrow{A})\,,
\label{htor}
\eeq
where
\beq
H_0^2=(c\overrightarrow{p}-e\overrightarrow{A})^2+
\hbar ce \overrightarrow{\Sigma}\cdot\overrightarrow{B}+
m^2c^4\,.
\eeq
The idea now is to consider the expansion of $\sqrt{H^2}$
not only in terms of the parameter $m$, but also in terms
of $S_\mu$. To perform this, we present the equation
(\ref{htor}) in the symmetric form
\beq
H^2&=&\frac{H_0^2}{2} \Big\{1+\frac{1}{H_0^2}
\left[2\eta_1mc^2
\overrightarrow{\Sigma}\cdot\overrightarrow{S}
+
2\eta_1\gamma_5S_0
\overrightarrow{\alpha}\cdot
(c\overrightarrow{p}-e\overrightarrow{A})\right]\Big\}
+
\cr
&+&
\Big\{1 + \left[2\eta_1mc^2
\overrightarrow{\Sigma}\cdot\overrightarrow{S}
+
2\eta_1\gamma_5S_0
\overrightarrow{\alpha}\cdot
(c\overrightarrow{p}-e\overrightarrow{A})\right]
\frac{1}{H_0^2}\Big\}
\frac{H_0^2}{2}
\,.
\label{htor2}
\eeq
This symmetrization is an important step of the procedure
which includes multiplication by $\beta$ in the equation
(\ref{ham0}). The next step is to extract the square root
of (\ref{htor2}). We expand the term $H_0^2$ in the power
series in $1/m$ (going to the second order in $1/m$) and
we obtain the same result of \cite {EK} which we call $H_0^{EK}$
\beq
H_o^{EK}=\sqrt{H_0^{2}}=mc^2+
\frac{(c\overrightarrow{p}-e\overrightarrow{A})^2}{2mc^2}
+\frac{\hbar e}{2mc}
\overrightarrow{\Sigma}\cdot\overrightarrow{B}\,.
\eeq

We also expand the term $1/H_0^2$ in the power series
in $1/m$ as well as the term in the brackets in (\ref{htor2})
in the power series of $S_\mu$, so that the result is of
the first order in $S_\mu$ and of the second order in $1/m$,
\beq
\sqrt{H^2}&=&H_0^{EK} + \eta_1
\overrightarrow{\Sigma}\cdot\overrightarrow{S}-
\frac{\eta_1}{2m^2c^4}
\overrightarrow{\Sigma}\cdot\overrightarrow{S}
(c\overrightarrow{p}-e\overrightarrow{A})^2 -
\cr
&-&
\frac{\hbar ce \eta_1}{2m^2c^4}
\overrightarrow{S}\cdot\overrightarrow{B} -
\frac{\eta_1}{mc^2}(c\overrightarrow{p}-e\overrightarrow{A})
\cdot
(S_0\overrightarrow{\Sigma} + i\gamma_5\be
\overrightarrow{S}\times\overrightarrow{\Sigma})\,.
\label{htransf2}
\eeq

The last term in (\ref{htransf2}) is odd, and using (\ref{tr})
we derive the final Hamiltonian for this case
\beq
H^{\prime \,\, tr}=\beta mc^2+
\beta\frac{(c\overrightarrow{p}-
e\overrightarrow{A}-
\eta_1 S_0 \overrightarrow{\Sigma}
-\be \eta_1 \overrightarrow{S}\times\overrightarrow{\Sigma} )^2}
{2mc^2}-
\frac{\be \eta_1}{2m^2c^4}(c\overrightarrow{p}-
e\overrightarrow{A})^2
\overrightarrow{\Sigma}\cdot\overrightarrow{S}
+
\nonumber
\eeq
\beq
+\beta\frac{\hbar ce}{2mc^2}
\overrightarrow{\Sigma}\cdot\overrightarrow{B}+
\beta \eta_1 \overrightarrow{\Sigma}\cdot\overrightarrow{B}-
\beta \frac{\hbar ce \eta_1}{2m^2c^4}
\overrightarrow{S}\cdot\overrightarrow{B}\,.
\label{htorstransf}
\eeq
Here, we used prime in $H$ in order to distinguish
between the Hamiltonians (\ref{ham}) and (\ref{htorstransf}).
For the Hamiltonian (\ref{htorstransf}) we apply the same
algorithm used between equations (\ref{expo}) and (\ref{eqexpo})
and then finally we obtain the Hamiltonian for the two-spinor
$\varphi$. The result can be expressed in the form
\beq
H^{\prime \,\, tr}_\varphi=
\left( 1-\frac{\overrightarrow{\Sigma}\cdot\overrightarrow{S}}
{2mc^2}\right) \, H^{tr}_\varphi \,
\left( 1-\frac{\overrightarrow{\Sigma}\cdot\overrightarrow{S}}
{2mc^2}\right)\,,
\label{ham2}
\eeq
where $H^{tr}_\varphi$ is given by the equation (\ref{ham}).
The next step is to derive the equations of motion using the
same procedure as was applied in \cite{BBS}. We perform the
canonical quantization of the theory introducing the operators
of coordinate $\,\hat{x}_i$, momenta $\,\hat{p}_i\,$ and spin
$\,\hat{\sigma}_i\,$ and implement the equal-time commutation
relations of the usual way. These operators yield the equations
of motion
\beq
i\hbar \frac{d\hat{x}_i}{dt}
= \left[\hat{x}_i, H \right]\quad,\quad
i\hbar \frac{d\hat{p}_i}{dt}
= \left[\hat{p}_i, H \right]\quad,\quad
i\hbar \frac{d\hat{\sigma}_i}{dt}
= \left[\hat{\sigma}_i, H \right]\,.
\label {commu}
\eeq
After the computation of the commutators in (\ref{commu}), we arrive
at the explicit form of the operator equations of motion. Now we can
omit all the terms which vanish when $\hbar \rightarrow 0$. Thus we
obtain the classical equations which can be interpreted as the
(quasi)classical  equations of motion for the particle in
an external torsion and electromagnetic fields. In this case
the equations of motion are
\beq
v_i=\frac{dx_i}{dt}=
\big(1-\frac{\eta_1}{mc^2}\overrightarrow{\sigma}
\cdot\overrightarrow{S}\big)
\,
\frac{1}{m}\,
(P_i-\frac{e}{c}A_i-\frac{\eta_1}{c} S_0 \sigma_i)
+
\frac{\eta_1}{mc}(\overrightarrow{\sigma}
\times\overrightarrow{S})_i\,,
\nonumber
\eeq
\beq
\frac{dp_i}{dt}=
\big(1-\frac{\eta_1}{mc^2}\overrightarrow{\sigma}
\cdot\overrightarrow{S}\big)\,
\frac{1}{m}\,
\left(p^j - \frac{e}{c}A^j - \frac{\eta_1}{c}
\sigma^jS_0 \right)\frac{e}{c}\frac{\pa A_j}{\pa x^i}
+\frac{\eta_1 e}{mc^2}\frac{\pa A^j}{\pa x^i}
(\overrightarrow{\sigma}\times\overrightarrow{S})_j\,,
\nonumber
\eeq
\beq
\frac{d\sigma_i}{dt} =
\left[ \overrightarrow{r}\times\overrightarrow{\sigma} \right]_i
\quad,\quad
\overrightarrow{r} =
\frac{2\eta_1}{\hbar}\left[ (1-\frac{v^2}{2c^2})
\overrightarrow{S}
+\overrightarrow{S}\times\frac{\overrightarrow{v}}{c}
- \frac{1}{c}\overrightarrow{v}S_0 \right]
+ \frac{e}{mc}\overrightarrow{B}\,.
\label{eqmotion3}
\eeq
Using the first two equations of (\ref{eqmotion3}), we write
\beq
m\frac{dv_i}{dt} &=& \left(
-\frac{e}{c}\frac{\pa A_i}{\pa t}
+\frac{e}{c}\left[ \overrightarrow{v}
\times\overrightarrow{B} \right]_i
\right)
\left(1-\frac{\eta_1}{2mc^2}\overrightarrow{\sigma}
\cdot\overrightarrow{S}\right)
-
\cr
&-&\frac{\eta_1}{c}\,\frac{d}{dt}\,(S_0 \, \sigma_i)
+\frac{\eta_1}{c}\,
\frac{d}{dt}\,(\overrightarrow{\sigma}\times\overrightarrow{S})_i
-\frac{\eta_1 \, v_i}{c^2}\,\frac{d}{dt}
(\overrightarrow{\sigma}\cdot\overrightarrow{S})
\,.
\label{eqmotion4}
\eeq

As compared to (\ref{eqmotion-ap}), the new terms in the
equation (\ref{eqmotion3}) are of the order $1/m^2$. The
equation (\ref{eqmotion4}) has the two important points.
The first is that the last term is of the order $1/m$ and
it was not present in equation (\ref{eqmotion2-ap}). This
result shows that the fact that we used only the parameter
$1/m$ in expansion of $H^2$ did not give us all the possible
linear terms with $S_\mu$ in the final Hamiltonian, as it
should be. The second point to note, is that the second term
in the equation (\ref{eqmotion4}) shows an interesting effect.
This equation is analogous to the Lorentz force acting on a
particle that interacts with an external electromagnetic field.
The term where $S_\mu$ appears can be seen as a correction for
this case. Thinking along these lines, this term shows an
explicit mixing between the torsion and the magnetic field.
One can imagine a situation when the magnetic field is strong
enough to compensate weakness of the spacetime torsion $S_\mu$
so that this term would affect particle's motion in a
notable way.
%%%%%%%%%%%%%%%%%%%%%%%%%%%%%%%%%%%%%%%%%%%%%%%%%%%%%%%%%%%
%%%%%%%%%%%%%%%%%%%%%%%%%%%%%%%%%%%%%%%%%%%%%%%%%%%%%%%%%%%%%%

\end{document}